\documentclass[aps,prx,floatfix,twocolumn]{revtex4-1}
\usepackage[english]{babel}

\makeatletter
\def\bbl@set@language#1{%
  \edef\languagename{%
    \ifnum\escapechar=\expandafter`\string#1\@empty
    \else\string#1\@empty\fi}%
  \@ifundefined{babel@language@alias@\languagename}{}{%
    \edef\languagename{\@nameuse{babel@language@alias@\languagename}}%
  }%
  \select@language{\languagename}%
  \expandafter\ifx\csname date\languagename\endcsname\relax\else
    \if@filesw
      \protected@write\@auxout{}{\string\select@language{\languagename}}%
      \bbl@for\bbl@tempa\BabelContentsFiles{%
        \addtocontents{\bbl@tempa}{\xstring\select@language{\languagename}}}%
      \bbl@usehooks{write}{}%
    \fi
  \fi}
\newcommand{\DeclareLanguageAlias}[2]{%
  \global\@namedef{babel@language@alias@#1}{#2}%
}
\makeatother

\DeclareLanguageAlias{en}{english}

\usepackage{graphicx}
\usepackage{epstopdf}
\usepackage{amsmath}
\usepackage{lineno}
\usepackage{comment}
\usepackage{graphicx, nicefrac}
\usepackage{float}
\usepackage[bookmarks=True,bookmarksopen=True]{hyperref} 

\usepackage{silence}
\usepackage[usenames,dvipsnames]{color}
\usepackage{colortbl}
\newcommand{\TV}[1]{} 

\hypersetup {
    unicode=false,          
    pdftoolbar=true,        
    pdfmenubar=true,        
    pdffitwindow=false,     
    pdfstartview={FitH},    
    pdftitle={Controlling sub-cycle instantaneous optical chirality in the photoionization of chiral molecules},    
    pdfauthor={Shaked Rozen},     
    pdfcreator={Shaked Rozen},   
    colorlinks=true,       
    linkcolor=BlueViolet,          
    citecolor=BlueViolet,        
    filecolor=Black,      
    urlcolor=Black           
}

\graphicspath{{./figures/}}  

\begin{document}


\title{Controlling sub-cycle optical chirality in the photoionization of chiral molecules}

\author{S. Rozen*$^{1}$,
A. Comby*$^{2}$,
E. Bloch$^{2}$,
S. Beauvarlet$^{2}$,
D. Descamps$^{2}$,
B. Fabre$^{2}$,
S. Petit$^{2}$,
V. Blanchet$^{2}$,
B. Pons$^{2}$,
N. Dudovich$^{1}$,
Y. Mairesse$^{2}$}
\bigskip

\affiliation{
$^1$ Weizmann Institute of Science, Rehovot, 76100, Israel\\
$^2$ Universit\'e de Bordeaux - CNRS - CEA, CELIA, UMR5107, F33405 Talence, France\\
*These authors contributed equally to this work.
}

\date{\today}



\begin{abstract}

{\bf
Controlling the polarization state of electromagnetic radiation enables the investigation of fundamental symmetry properties of matter through chiroptical processes. Over the past decades, many strategies have been developed to reveal structural or dynamical information about chiral molecules with high sensitivity, from the microwave to the extreme ultraviolet range. Most schemes employ circularly or elliptically polarized radiation, and more sophisticated configurations involve, for instance, light pulses with time-varying polarization states. All these schemes share a common property -- the polarization state of light is always considered as constant over one optical cycle. In this study, we zoom into the optical cycle in order to resolve and control a subcyle attosecond chiroptical process. We engineer an electric field whose instantaneous chirality can be controlled within the optical cycle, by combining two phase-locked orthogonally polarized fundamental and second harmonic fields. While the composite field has zero net ellipticity, it shows an instantaneous optical chirality which can be controlled via the two-color delay. We theoretically and experimentally investigate the photoionization of chiral molecules with this controlled chiral field. We find that electrons are preferentially ejected forward or backward relative to the laser propagation direction depending on the molecular handedness, similarly to the well-established photoelectron circular dichroism process. However, since the instantaneous chirality switches sign from one half cycle to the next, electrons ionized from two consecutive half cycles of the laser show opposite forward/backward asymmetries. This chiral signal, termed here as ESCARGOT (Enantiosensitive Sub-Cycle Antisymmetric Response Gated by electric-field rOTation), provides a unique insight into the influence of instantaneous chirality in the dynamical photoionization process. More generally, our results demonstrate the important role of sub-cycle polarization shaping of electric fields, as a new route to study and manipulate chiroptical processes.

}
\end{abstract}


\maketitle

\section{Introduction}
Since its discovery by Biot in the XIX$^{th}$ century, circularly polarized light (CPL) has served as the tool of choice to reveal and investigate the properties of chiral molecules \cite{SpectroscopyBook2012}. The polarization state of light is quantified by the degree of ellipticity -- the aspect ratio of the polarization ellipse, described by the electric field over one optical period. As such, the ellipticity is a cycle-averaged quantity. Recent progress in attosecond spectroscopy enables physicists to observe effects taking place within an optical cycle and sample its oscillation with attosecond precision \cite{goulielmakis04,kim13}. Clever designs have been developed, demonstrating the ability to record the complete temporal evolution of complex vectorial fields, whose polarization state varies on extremely short timescales \cite{boge14,carpeggianni17}. Shaping the sub-cycle polarization state of light offers an elegant way to measure and control strong field processes and to resolve structural and temporal properties of electronic wavefunctions in atoms and molecules \cite{kitzler05,shafir09,shafir12,zhang14,kfir15,richter15,smirnova15,ferre16,wurzler17,neufeld18,baykusheva18,li19}.

The interaction of light whose ellipticity varies along an optical cycle, with chiral matter, raises fundamental questions. Is the variation of the direction of the electric field over a few hundreds of attoseconds sufficient to induce dichroism effects? Can sub-cycle vectorial fields be used to resolve attosecond-scale chiroptical processes? Answering these questions raises three fundamental challenges. First, we should find a proper mathematical description of the instantaneous chirality of light. Next, we have to establish a control scheme to manipulate this chirality. Finally, we must find a physical observable that is sensitive to manipulations of the sub-cycle chirality.

In order to describe the instantaneous chirality of an electromagnetic field within the optical cycle limit, we have to go beyond the cycle averaged definition of ellipticity. The \textit{instantaneous ellipticity} (IE) is defined according to the sub-cycle evolution of two orthogonally polarized fields, dictated by their temporal dephasing within the optical cycle \cite{Strelkov2004}. The IE reflects the temporal evolution of the ellipticity if the central frequency of an ultrashort pulse is much larger than its spectral bandwidth, as is typically the case in coherent control experiments \cite{Misawa2016}. However, this measure does not capture the instantaneous change in the rotation direction of the electromagnetic field, which is directly related to the chirality. A deeper insight into the chiral properties of light is described by a time-even pseudoscalar characterizing an electromagnetic field called ``zilch'', introduced by Lipkin in 1964 \cite{lipkin1964} and renamed optical chirality in the early 2000's by Tang and Cohen \cite{tang10,tang11}. This concept was mostly used in the spatial domain -- spatially shaping light beams enables optical chiralities above 1 to be produced, resulting in enhanced circular dichroism signals \cite{tang11}. In the temporal domain, the optical chirality was recently shown to describe the rotational velocity of the light electric field, and used to theoretically investigate chiral-light / achiral matter interaction \cite{neufeld18}.

\begin{figure}[!t]
	\begin{center}
		\centering{\includegraphics*[width=\columnwidth]{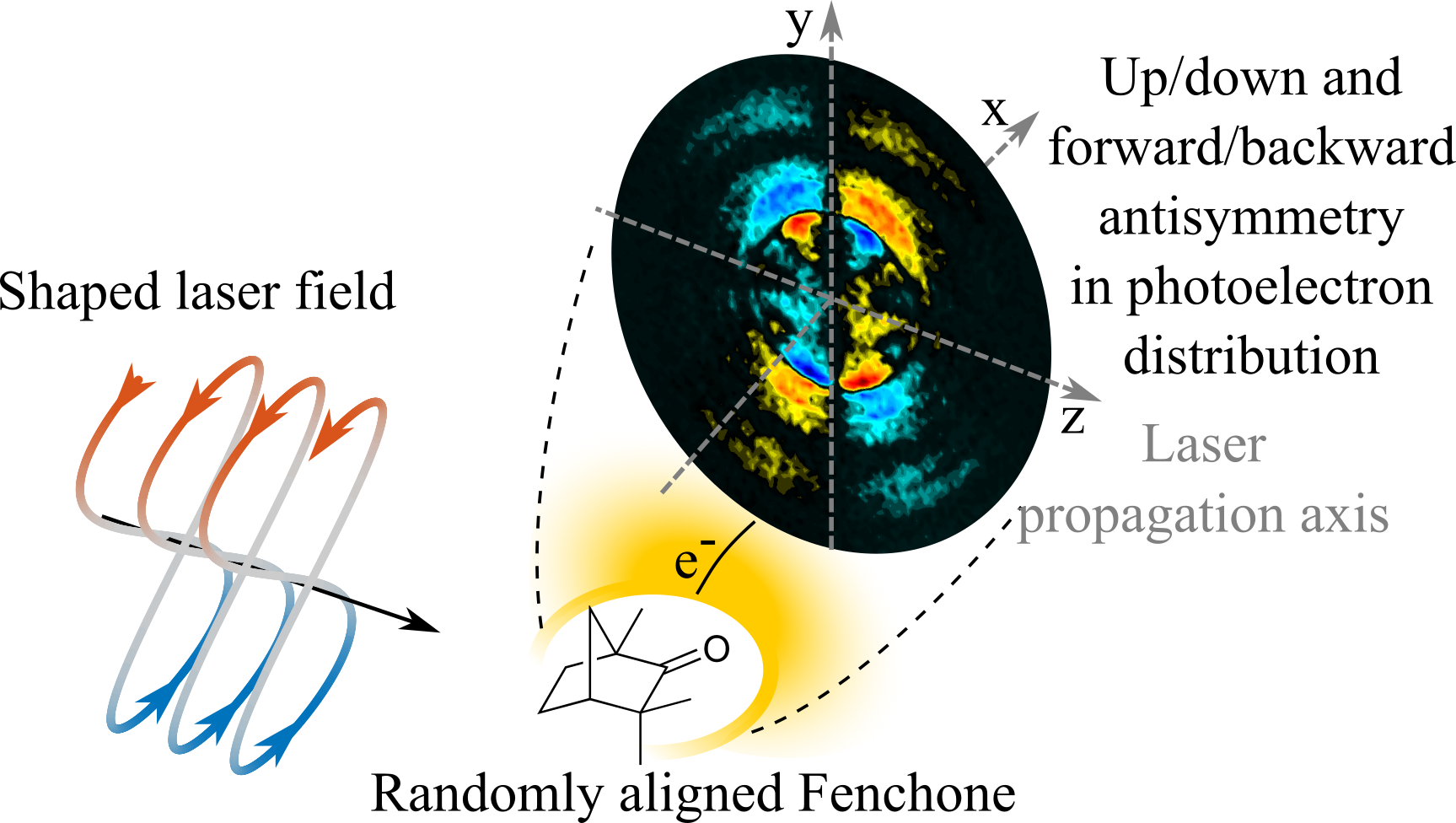}}
		\caption{Principle of the experiment: the shaped electric field is focused into a jet of randomly aligned enantiopure fenchone molecules, in the interaction zone of a VMI spectrometer. The VMI records the photoelectron momentum distribution, whose forward/backward antisymmetric component is shown in the figure. This component shows a clear up/down antisymmetry. }
		\label{Concept}
	\end{center}
\end{figure}

In this work, we manipulate the instantaneous optical chirality of a laser field by superimposing a fundamental component and its phase-locked orthogonally-polarized second harmonic. This field configuration was recently proposed theoretically as a way to produce asymmetric responses in the weak-field photoionization of chiral molecules \cite{Baumert2018}. Our work investigates both theoretically and experimentally these asymmetric responses in the multiphoton and strong field regimes. Due to symmetry, the ionizing field has a zero net chirality: it describes opposite rotations in two consecutive half cycles of the fundamental radiation.  The time-integrated interaction of this field with a chiral molecule will thus not discriminate between two enantiomers. However, this field has nonzero instantaneous optical chirality, whose temporal evolution can be controlled by manipulating the two-color delay. In order to resolve the sub-cycle chiral nature of this scheme, we photoionize chiral molecules and detect the angular distribution of the ejected photoelectrons. Circularly polarized light is known to preferentially eject photoelectrons from chiral molecules in the forward or backward direction, depending on the light's and molecule's handedness. This mechanism leads to a very strong chiroptical effect, referred to as photoelectron circular dichroism (PECD)  \cite{Ritchie1976,powis00,bowering01,nahon15}. 
As shown in Figure~\ref{Concept}, photoionization with our engineered composite field creates a significant forward/backward asymmetry in the collected photoelectrons distribution. Moreover the forward antisymetry in the  upper hemisphere corresponds to a backward antisymety in the lower hemisphere, revealing that electrons emitted from two consecutive half cycles of the composite field experience opposite light handedness. This forward-backward/up-down antisymetry varies drastically with the two-color delay, allowing for a control of the subcycle chiral interaction.
We refer to this scheme as ESCARGOT (snail in French), standing for Enantiosensitive Sub-Cycle Antisymmetric Response Gated by electric-field rOTation.  In this work we present theoretical and experimental data for both the multiphoton and strong field ionization regimes. Since the features shown in the multiphoton regime are very complex, we present a simple model explaining the origin of the signal in the strong field regime, and leave the more complex analysis of the multiphoton signal to a later stage. A qualitative classical analysis of the departing electron trajectories shows intuitively how the measured asymmetry patterns stem from the accumulation of the instantaneous field chirality over the first few hundreds of attoseconds the electron takes to leave the inner chiral molecular region. These results thus demonstrate the sensitivity of photoionization to the sub-cycle instantaneous chirality and, more generally, reveal the important role played by this physical quantity in chiral light-matter interaction.

\section{Controlling sub-cycle optical chirality}

\begin{figure}[!t]
\begin{center}
  \centering{\includegraphics*[width=\columnwidth]{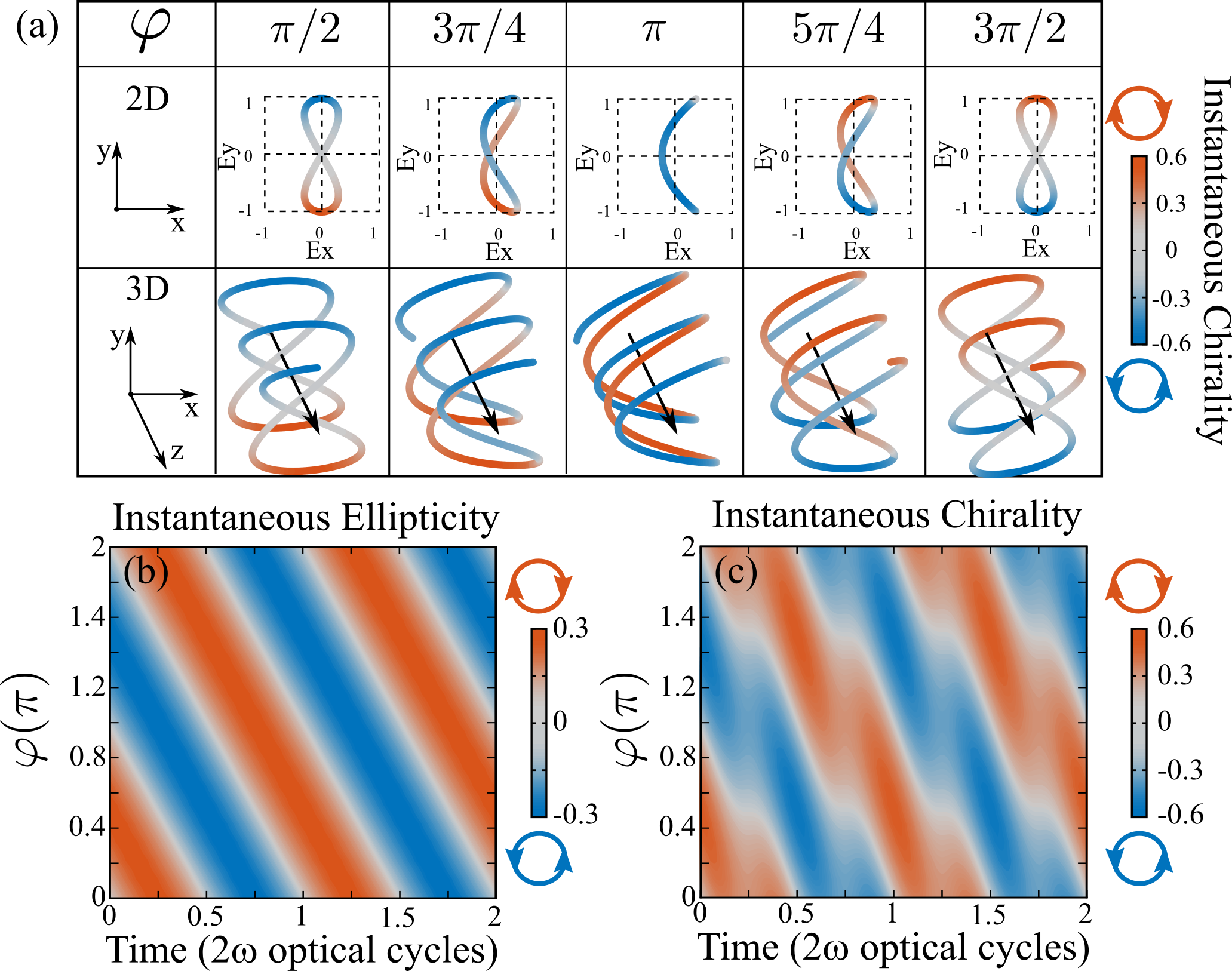}}
  \caption{Schematic description of the sub-cycle instantaneous ellipticity and optical chirality of the pulse. (a) 2D and 3D representations of the shape of a composite electric field obtained by combining a fundamental component at $\omega$ polarized along $y$ and a $2\omega$ component along $x$, with 10$\%$ intensity ratio, for different relative phases $\varphi$ between the two components. The color indicates the instantaneous chirality at each instant in the cycle. (b) Instantaneous ellipticity and (c) Instantaneous chirality of the composite field along one laser cycle, as a function of $\varphi$. The instantaneous ellipticity and chirality are defined in the main text (eq. \ref{epsilon},\ref{eq:1}). The instantaneous chirality is normalized relative to the chirality of a left circularly polarized field. The normalization of the instantaneous ellipticity is implied in its definition. Note that in 2D representation of the $0$ and $\pi$ phases, we can only see the first half cycle of the laser field. The second half cycle is not visible since it traces exactly the same route as the first half, only in the reverse direction, which causes the chirality of the second half to be opposite to that of the first half, as seen in the 3D representations.}
  \label{EllipticityChirality}
\end{center}
\end{figure}

To control the sub-cycle optical chirality, we use two phase-locked orthogonally polarized laser fields: the superposition of a fundamental field and its second harmonic. This superposition can produce a variety of shapes of the vectorial electric field, dictated by the delay between the two fields. The two-color field is described as:
\begin{equation}
 {\bf E}(t)=E_0 \cos(\omega t){\bf \hat{y}}+r E_0 \cos(2\omega t+\varphi){\bf \hat{x}}
 \end{equation}
  where $\omega$ is the fundamental frequency, $r$ is the ratio between field amplitudes, and $\varphi$ is the relative phase between the two components. In our experiments and calculations we set the relative amplitudes to be $r\approx0.3$. We illustrate in Figure~\ref{EllipticityChirality}(a) how changing the two-color phase $\varphi$ influences the shape of the composite ($\omega$,$2\omega$)-field in the transverse ($x,y$)-plane. For instance, the field exhibits a 8-shape 
aligned along $y$ for $\varphi=\pi/2$. Such a field can be intuitively understood as successive left and right elliptically polarized half-cycle components producing photoelectrons in the upper and lower hemispheres of the VMI detector, respectively. For $\varphi=\pi$, the field follows a C-shape which has no direct analogy with any common polarization pattern. 

The net ellipticity and the net chirality of the composite field are zero for all two-color phases. A deeper insight into the $\varphi$-dependent properties of 
the field can be obtained by looking into the subcycle variation of its vectorial properties. First we focus on the evolution of the instantaneous ellipticity within the optical cycle, defined as \cite{boge14}:   
\begin{align} \label{epsilon}
\epsilon(t)=\tan \left[ \frac{1}{2} \arcsin \left( \frac{2E_x(t)E_y(t)\sin(\phi)}{E_x^2(t)+E_y^2(t)} \right) \right]
\end{align}
where $E_x$ and $E_y$ are the ${\bf \hat{x}}$ and ${\bf \hat{y}}$ components of the total electric field ${\bf E}$. 
This instantaneous ellipticity is displayed in Figure~\ref{EllipticityChirality}(b) as a function of the two-color delay. It modulates with 
$2\omega$ frequency, reflecting the dephasing of the orthogonally polarized second harmonic field with respect to the fundamental cycle. The instantaneous ellipticity shows the same temporal shape, simply shifted in time as the relative phase is scanned. This quantity is thus unable to describe the strong changes in the instantaneous behavior of the electric field as a function of $\varphi$. An alternative description of the vectorial properties of the field consists of its \textit{instantaneous chirality}, defined as ~\cite{neufeld18}: 
\begin{align} \label{eq:1}
C(t)=\frac{\epsilon_0}{2 c_0}(E_y(t)\partial_t E_x(t)-E_x(t)\partial_t E_y(t))\\\nonumber
+\frac{1}{2 c_0 \mu_0}(B_y(t)\partial_t B_x(t)-B_x(t)\partial_t B_y(t))
\end{align}
where $c_0$ is the speed of light in vacuum, $B_x$ and $B_y$ are the ${\bf \hat{x}}$ and ${\bf \hat{y}}$ components of the magnetic field ${\bf B}$ associated to ${\bf E}$. Here we ignore the contribution of the magnetic field, since its effect on chiral photoionization is negligible compared to the electric dipole component associated with PECD. $C(t)$ is further shown in Figure~\ref{EllipticityChirality}(c) as a function of the two-color delay. The magnitude of the instantaneous chirality is superimposed on the two- and three-dimensional representations of the field shapes in Figure~\ref{EllipticityChirality}(a). The instantaneous chirality shows a more complex behavior as a function of the relative phase $\varphi$ than the instantaneous ellipticity. 
When $\varphi=\pi/2$, both quantities evolve similarly within the cycle: $\epsilon(t)$ and $C(t)$ are both maximum on top of the 8-shape, where the amplitude of the fundamental field is the largest, and they both reach their minimum value at the bottom of the 8. On the other hand, when $\varphi =\pi$, the instantaneous chirality is maximal when the instantaneous ellipticity is close to zero. The color-coding of the shape of the electric field presented in Figure~\ref{EllipticityChirality}(a) reveals the origin of the complex behavior of the instantaneous chirality: it maps the rotational velocity of the electric field \cite{neufeld18}.

\section{Photoelectron circular dichroism}
The ability to manipulate the sub-cycle vectorial properties of the field raises the following question : can we observe a chiroptical signal induced by the instantaneous chirality of the field while the net chirality of the light is zero over the optical period? In most cases, chiroptical processes are induced by bound electron dynamics, mainly driven by weak electric-quadrupole or magnetic-dipole interactions. Purely electric dipole chiroptical phenomena have recently been demonstrated, greatly improving the sensitivity of measurements \cite{Ritchie1976,patterson13,BeaulieuNatPhys2018}. For instance, when CPL is applied to photoionize chiral molecules, it provides an extremely sensitive probe of chirality via the PECD effect. This process, first predicted several decades ago ~\cite{Ritchie1976}, was observed in the single photon ~\cite{Heinzmann2001,nahon15}, multiphoton \cite{Baumert2012,lehmann13} and strong-field \cite{Beaulieu2016} ionization regimes.

PECD results from the scattering dynamics of the outgoing electrons in the chiral molecular potential, under the influence of the circularly polarized electric field.  A common property shared by all PECD experiments up to now is the constant polarization state of the laser field -- may it be circularly \cite{nahon15} or elliptically \cite{lux15,Comby2018} polarized: the PECD process is an accumulation over the few optical cycles during which the electron leaves the ionic core. Attosecond metrology opened a new path in both measurement and control of the photoionization dynamics in simple systems as rare gas atoms \cite{LHuillier2011,Dahlstorm2014,Keller2017} or diatomics \cite{Haessler2009,vos_orientation-dependent_2018}, allowing their probing on a sub-cycle level. Transferring basic attosecond metrology approaches to PECD measurements holds the potential of revealing the attosecond scale dynamical properties of chiral photoionization. A first step in that direction was recently investigated with the measurement of the photoionization delays of electrons emitted from chiral molecules: it was found that electrons ejected forward and backward could be delayed by a few tens of attoseconds \cite{BeaulieuScience2017}.

Here, we aim at detecting the sensitivity of chiral photoionization to instantaneous chirality (Figure~\ref{Concept}). Let us for instance consider the case where the field describes an 8 shape ($\varphi=\pi/2$). Since the net chirality of the ionizing light is zero, we should not expect any overall forward-backward asymmetry in the photoelectron angular distribution -- the PECD should be zero. However the electric field that we engineer has an instantaneous chirality, whose effect can be resolved by detecting the electrons ejected up and down in the photoionization process: in each of the upper and lower hemispheres of the VMI detector, a forward-backward asymmetry could emerge in the photoelectron angular distribution, with respect to the laser axis propagation, but the asymmetries in the upper and lower hemispheres should be opposite. In other words, the chiral-sensitive signal should be up/down and forward/backward antisymmetric. A recent theoretical investigation of chiral photoionization by a superposition of orthogonally polarized fundamental and second harmonic pulses predicted the existence of such asymmetries, as the result of quantum interference between photoionization pathways in a coherent-control scheme \cite{Baumert2018}. These calculations also predicted that the asymmetry vanishes, for the investigated system in the perturbative regime, when the relative phase between the two components is $\varphi=0$ (C-shaped field). In the present work, we focus on the extremely non-linear (high-order) multiphoton and strong field ionization regimes.

\section{Toy-model quantum calculations}
\begin{figure*}[!t]
	\begin{center}
		\centering{\includegraphics*[width=2\columnwidth]{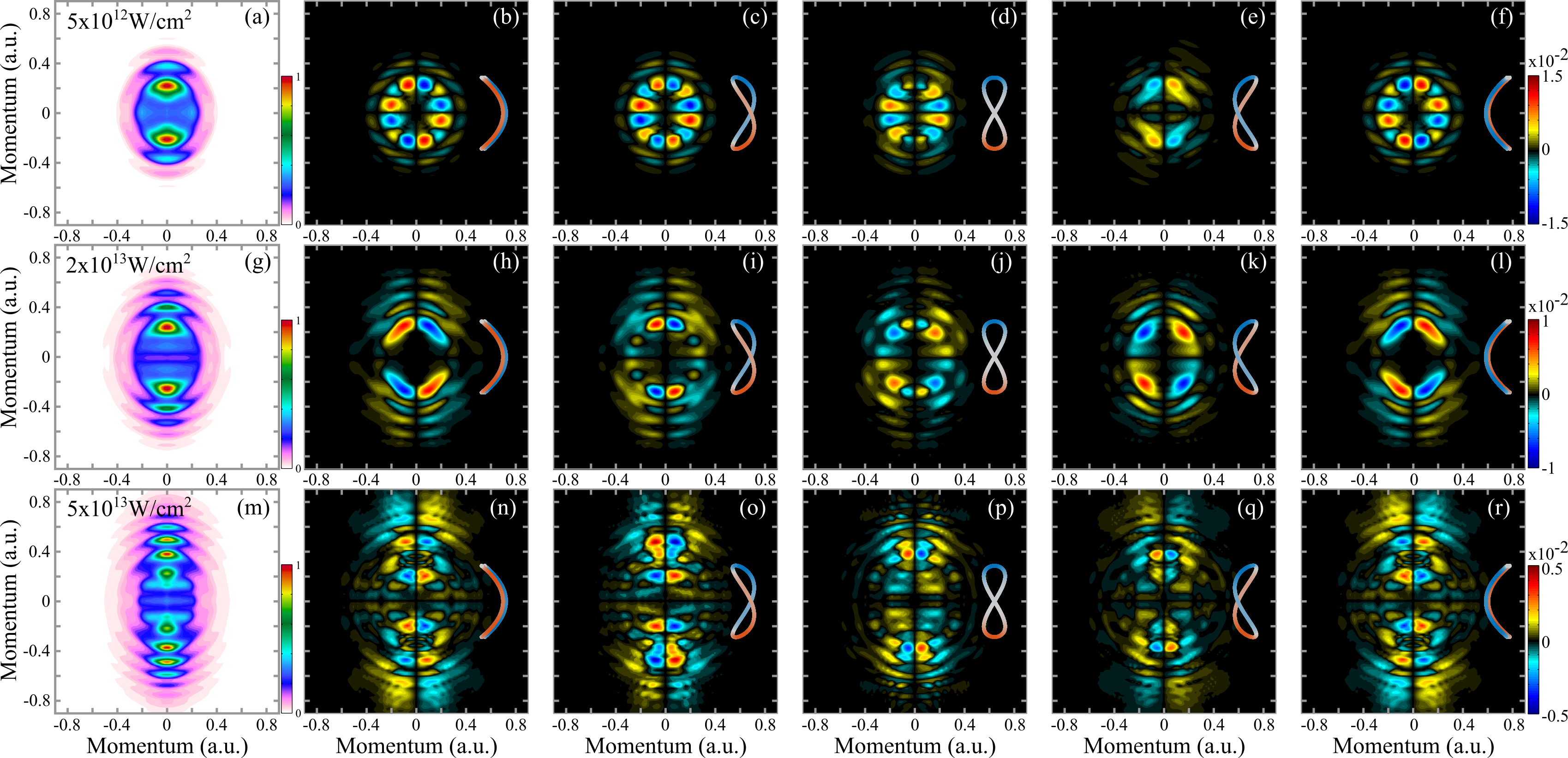}}
		\caption{{TDSE calculations of photoionization of a toy model chiral molecule by a composite bichromatic field. (a,g,m) Projections in the y-z plane of the symmetric part of the photoelectron spectrum, at $I=5\times 10^{12}$ (a), $2\times 10^{13}$ (g) and $5\times 10^{13}$ W/cm$^2$ (m) . The Keldysh parameter for these calculations is $\gamma=3.9,\gamma=1.9,\gamma=1.2$ respectively. (b-f),(h-l),(n-r) Corresponding projections of the ESCARGOT signal, defined as the antisymmetric part of the photoelectron distribution with respect to the propagation axis z and fundamental laser polarization axis y. The relative phase between the two components of the laser field are $\varphi=0$, $\pi/4$, $\pi/2$, $3\pi/4$ and $\pi$ from left to right. The shape of the two-color field, colored according to the instantaneous chirality, is shown next to each calculation.
		}}
		
		\label{figTH}
	\end{center}
\end{figure*}

First we perform quantum calculations by solving the time-dependent Schr\"odinger equation (TDSE) describing the interaction of a bichromatic field with a toy model 
chiral molecule. Our system consists of an electron evolving in the combined Coulomb fields of four nuclei with charges
$Z_1=-1.9$ and $Z_{2-4}=0.9$ a.u., respectively located at ${\bf R_1}={\bf 0}$, ${\bf R_1}= {\bf \hat{x}}$, ${\bf R_2}=2{\bf \hat{y}}$ and 
${\bf R_3}=3{\bf \hat{z}}$. Atomic units are used throughout this section unless otherwise stated. The charges and internuclear distances have been chosen to obtain (i)
 an ionization potential $I_P=8.98$ eV close to that of fenchone and camphor, the two chiral molecules experimentally studied in this work, and (ii) a PECD around $\sim 2\%$ when ionizing with circularly polarized light ($C(t)=\pm 1$). 

We work within a spectral approach where the bound and continuum eigenstates $\phi_{E_i}$ of the molecule are obtained by diagonalizing the field-free Hamiltonian 
$H_0=-\frac{1}{2}\nabla^2-\sum_{i=1}^{4}{\frac{Z_i}{|{\bf r}-{\bf R_i}|}}$, where ${\bf r}$ is the electronic coordinate vector. Details on the basis of
primitive functions employed to diagonalize $H_0$ are given in the Methods section.  We then solve the TDSE: 
\begin{equation} \label{TDSE}
\left( H_0+V({\bf \hat{R}},t)-i\frac{\partial}{\partial t} \right) \Psi( {\bf \hat{R}};{\bf r},t)=0
\end{equation}
in the molecular frame by expanding the total wavefunction $\Psi({\bf \hat{R}};{\bf r},t)$ onto the molecular eigenstates according to
$\Psi({\bf \hat{R}};{\bf r},t)=\sum_{i}{a_i({\bf \hat{R}};t) \phi_{E_i}({\bf r}) \exp(-i E_i t)}$, subject to the initial conditions
$a_i({\bf \hat{R}};0)=\delta_{i0}$ where the index 0 stands for the ground state. Inserting the eigenstate expansion into eq. (\ref{TDSE}) 
yields a system of coupled differential equations for the expansion coefficients $a_i({\bf \hat{R}};t)$ which is solved using standard
numerical techniques (see Methods section). 
The orientation of the molecular frame with respect to the lab frame is described by ${\bf \hat{R}}$, expressed in terms of the Euler angles $\left( \alpha, \beta,\gamma \right)$. $V({\bf \hat{R}},t)$ is the laser-molecule interaction potential :
$V({\bf \hat{R}},t)={\bf r\cdot E(t)}$ or $V({\bf \hat{R}},t)=-i{\bf A(t)\cdot \nabla}$ in the length and velocity gauges, respectively. 
Both gauges lead to the same dynamical results if we employ a large underlying $\phi_{E_i}$ basis. However, we find that similarly to \cite{Cormier1996}, the 
velocity gauge fastens the convergence of the calculations. 
The field ${\text {\bf E}}(t)$ potential vector ${\text {\bf A}}(t)$, defined in the lab frame, are passively rotated to the molecular one 
where the TDSE (\ref{TDSE}) is solved, using the Euler rotation matrix ${\cal R}(\hat{{\text {\bf R}}})$ \cite{Goldstein}. 

We consider a flat-shape 4-cycle fundamental pulse with $\lambda=800$ nm with one $\omega$-cycle ascending and descending ramps. These parameters are far from the experimental conditions described in the next section, where a 1030 nm pulse with 130 fs duration was used. Our  objective here is to explore qualitatively the potential of sub-cycle shaped fields to produce enantiospecific antisymmetric signals in the photoionization of chiral molecules, in the high-order multiphoton and strong-field regimes, without trying to quantitatively reproduce the experiment. In that context a 800 nm field offers a good compromise between computation duration (which increases nonlinearly with $\lambda$) and the access to the tunnel-ionization regime (which scales as $1/\lambda^2$). 

At final time $t_f$, we extract the ionized part of the total 
wavefunction in the molecular frame under the restriction $E_i>0$, {\em i.e.}
\begin{align}\label{psi_ion}
\Psi^{ion}({\bf \hat{R}};{\bf r},t_f)=\sum_{i,E_i>0}{a_i({\bf \hat{R}};t_f) \phi_{E_i}({\bf r}) {\text e}^{-i E_i t_f}}.
\end{align}
We show in the Methods section how we rotate this wavefunction into the lab frame and how we transform it to momentum space, yielding: $\Psi^{ion}_{lab}({\bf \hat{R}};{\bf p},t_f)$ 
where ${\bf p}$ is the momentum vector of the electron. We obtain the momentum density of ionized electrons from the 
sample of randomly oriented molecules as
\begin{align}\label{rho_p}
\rho({\bf p})=\int{d{\bf \hat{R}} |\Psi^{ion}_{lab}({\bf \hat{R}};{\bf p},t_f)|^2}.
\end{align}
The integral is evaluated as a numerical quadrature over countable molecular orientations with Euler angular spacing 
$\Delta \alpha = \Delta \beta = \Delta \gamma = \pi /4$ (see Methods section for convergence). Finally, we integrate the total density $\rho({\bf p})$ over $p_x$ to mimic detection by a velocity map
imaging (VMI) spectrometer with time-of-flight along ${\bf \hat{x}}$, yielding
\begin{align}\label{rho_{VMI}}
\rho_{VMI}(p_y,p_z)=\int{dp_x \rho({\bf p})}. 
\end{align}

Our calculations explore the chirosensitive response for laser intensities ranging from the multiphoton to the strong field regimes. In these regimes, intermediate resonant states are not expected to play a major role, contrarily to the coherent control regime explored in \cite{Baumert2018}. To analyze the results, we decompose the photoelectron angular distributions $\rho_{VMI}(p_y,p_z)$ in symmetric 
$\rho^{sym}_{VMI}(p_y,p_z)=[\rho_{VMI}(p_y,p_z)+\rho_{VMI}(p_y,-p_z)]/2$ and antisymmetric 
$\rho^{anti}_{VMI}(p_y,p_z)=[\rho_{VMI}(p_y,p_z)-\rho_{VMI}(p_y,-p_z)]/2$ parts along the light propagation axis $z$ \cite{Baumert2018}. 
However the calculations employ short pulses whose envelope shape leads to symmetry breaking of the field in the $y$-direction (carrier-envelop phase effects). 
This results in not perfectly symmetric $\rho^{sym}_{VMI}$ and antisymmetric $\rho^{anti}_{VMI}$ distributions. 
We therefore correct the consequences of the short pulse duration on electron distributions by defining: 
\begin{eqnarray}
\rho^{sym}_{VMI}(p_y,p_z)=[ \rho_{VMI}(p_y,p_z)+ \rho_{VMI}(p_y,-p_z)\\
+\rho_{VMI}(-p_y,p_z)+ \rho_{VMI}(-p_y,-p_z)] /4 \nonumber
\end{eqnarray}
and 
\begin{eqnarray}\label{rho_anti}
\rho^{anti}_{VMI}(p_y,p_z)=[ \rho_{VMI}(p_y,p_z)- \rho_{VMI}(p_y,-p_z)\\
-\rho_{VMI}(-p_y,p_z) + \rho_{VMI}(-p_y,-p_z)] /4 \nonumber
\end{eqnarray}
whose sum yields the average $[\rho_{VMI}(p_y,p_z)+\rho_{VMI}(-p_y,-p_z)]/2$ as expected. The influence of the antisymmetrization procedure is 
shown in the Methods section. Importantly, $\rho^{anti}_{VMI}$ is further normalized to the maximum of $\rho^{sym}_{VMI}$ in order to quantify the asymmetry 
as a fraction of the maximum electron count.

Figure \ref{figTH}(a,g,m) shows $\rho^{sym}_{VMI}(p_y,p_z)$ for 800 nm-laser intensities $I=5\times 10^{12}$, $2\times 10^{13}$ and $5\times 10^{13}$ W/cm$^2$ and $r=0.3$. We observe a clear transition from a low-intensity multiphoton ionization regime, in which only a few above-threshold ionization (ATI \cite{Agostini1979}) peaks are present, to a strong-field ionization regime showing many ATI peaks and a sharpening of the photoelectron distribution around the laser polarization plane. 

The chiral-sensitive part of the signal, to which we refer as ESCARGOT, is represented by $\rho^{anti}_{VMI}(p_y,p_z)$. Figure~\ref{figTH}(b-f,h-l,n-r) presents the ESCARGOT signal as a function of the two-color delay for the three laser intensities. A significant ESCARGOT signal shows up for {\em all} two-color delays and in {\em all} ionization regimes. In the multiphoton regime, the ESCARGOT signal is highly structured angularly, with distinct contributions of alternating signs. This is the signature of the large number of photons absorbed to ionize the molecule, which is known in multiphoton ionization induced by circularly polarized light \cite{Beaulieu2016,Baumert2012}. The asymmetry reaches $\sim 1.5\%$, which is weaker than the conventional PECD obtained using monochromatic circular light. As we increase the laser intensity, we enter the strong-field regime where the ESCARGOT signal decreases down to $\sim 0.5 \%$, and sharpens about the polarization plane, consistently with the behavior of the whole photoelectron distribution. 
Traces of ring structures inherent to ATI peaks are still noticeable in the symmetric part of the signal. These stuctures
are more prominent in the ESCARGOT pictures where the asymmetry is found to alternate sign along the rings associated to low electron energies. 
The magnitude and angular dependence of ring asymmetries depends on the two-color phase, which points towards a strong interplay of 
field-induced ionization dynamics and underlying core chirality of the molecular target. The ESCARGOT pattern is simpler in the high electron energy 
range where the ATI peaks present the same asymmetry, centred about the $\omega$-2$\omega$ polarization plane. 

Generally, one remarkable feature in our results is that we observe a significant ESCARGOT signal for all shapes of the electric field, even with the C-shaped field, in all ionization regimes investigated here. 
This could seem counter-intuitive, since the C-shaped field exhibits no preferable rotation direction in the upper and lower hemispheres \cite{Baumert2018}. Indeed, Demekhin \textit{et al.} observed no asymmetry using a C-shaped field in a single vs two photon ionization scheme, in which the ionization can be considered as a temporally continuous process. By contrast, in our case the strong multiphoton character of the process leads to the temporal confinement of the ionization around the maxima of the laser field. The ionization has to be treated as a dynamical phenomenon. For instance in the strong-field limit, the electron dynamics will be driven by the temporal evolution of the vector potential  \cite{Corkum1994}. While the C shaped field rotates in the same direction in the lower and upper hemispheres, its time derivative -- the vector potential -- describes an 8 shape with opposite rotations, imposing an up-down antisymmetry in the chiral part of the photoelectron distribution. Our results indicate that at laser intensities much below the strong field limit, the ionization dynamics remains sensitive to the temporal evolution of the electric field, since we observe ESCARGOT signals for all field shapes. The high sensitivity to the electric field evolution is further illustrated by the strong difference between the ESCARGOT signals obtained with very similar shapes of the electric field, at $\varphi=\pi/4$ and $\varphi=3\pi/4$.

\section{Experimental results}

\begin{figure*}[!t]
	\begin{center}
		\centering{\includegraphics*[width=2\columnwidth]{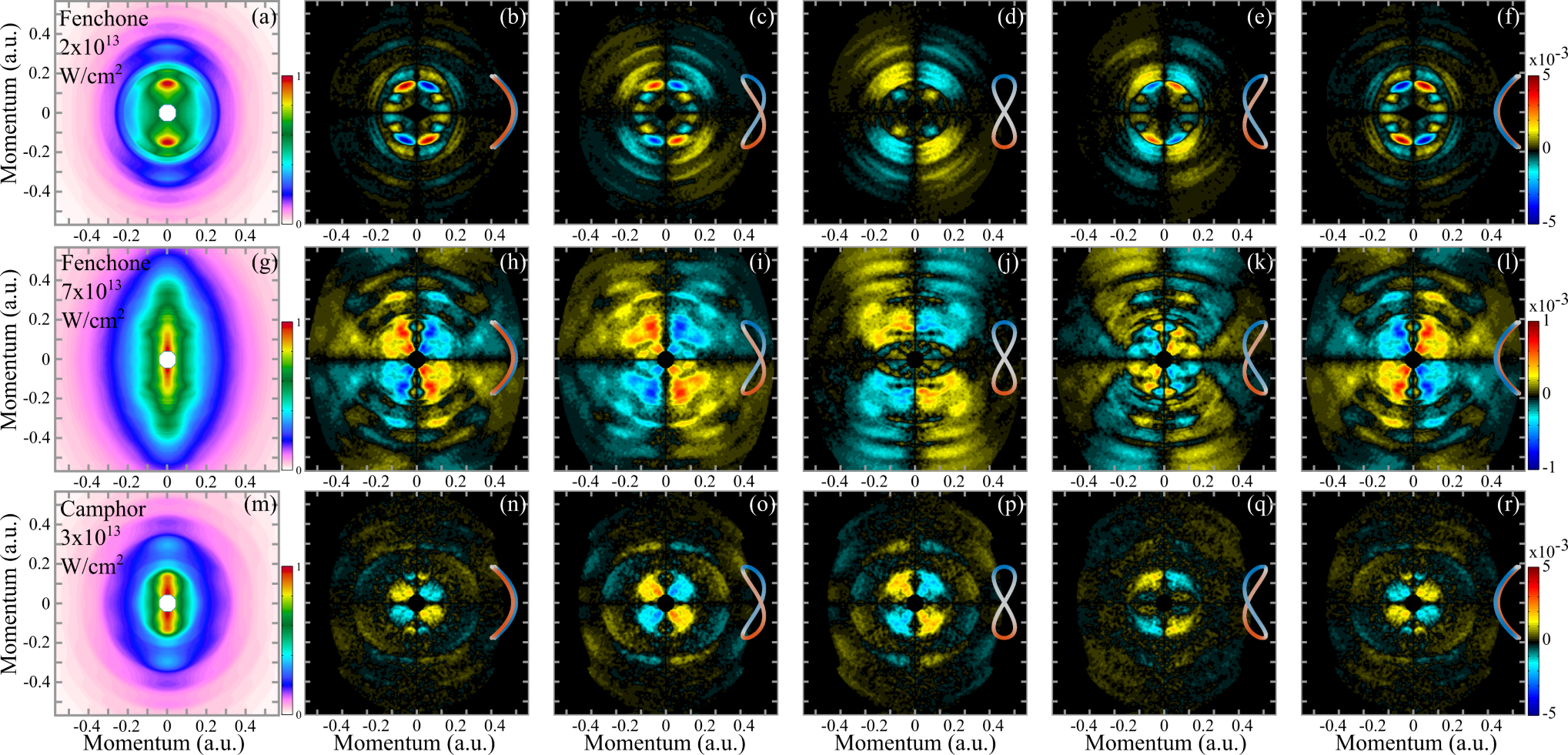}}
		\caption{{Experimental data. (a,g,m)  Normalized projections in the y-z plane of the the photoelectron angular distribution for (a) Fenchone(+) with a fundamental field of $2\times10^{13}$ W.cm$^{-2}$, (g) Fenchone(+) at $7\times10^{13}$ W.cm$^{-2}$ and (m) Camphor(+) at $3\times 10^{13}$ W.cm$^{-2}$. The Keldysh parameter in these data sets is $\gamma=1.5$ for (a),(m) and $\gamma=0.8$ for (g). (b-f),(h-l),(n-r) Evolution of the ESCARGOT signal as a function of the relative phase $\varphi$ between the two components of the ionizing field ($\varphi=0$, $\pi/4$, $\pi/2$, $3\pi/4$ and $\pi$ from left to right).  The shape of the two-color field, colored according to the instantaneous chirality, is shown next to each image. The ESCARGOT signal is normalized to the maximum of the photoelectron spectrum in (a,g,m). The data is presented after up-down antisymmetrization (see methods).
		}}
		
		\label{FigXP}
	\end{center}
\end{figure*}

In the following step we perform an experimental study of the ESCARGOT signal in (+)-Fenchone and (+)-Camphor molecules. The experiment was conducted using the Blast Beat laser system at CELIA (dual Tangerine Short Pulse, Amplitude Systems). We used a 50 W beam of 130 fs pulses at 1030 nm ($\omega$), at a repetition rate of 750 kHz. The beam was split into two arms, balanced by a polarizing beamsplitter associated with a half waveplate. One beam was frequency doubled in a 1 mm type-I beta barium borate (BBO) crystal to obtain a 515 nm beam at 2 W (2$\omega$), while the second arm remained at the fundamental frequency. The polarization state of the 515 nm arm was rotated by 90$^\circ$ using a half wave plate. The two arms were recombined by a dichroic mirror, and the relative intensities were adjusted to keep an intensity ratio of $\nicefrac{I_{2\omega}}{I_{\omega}}=0.1$, in order to maintain a field ratio of $r=0.3$. The relative delay was controlled by a pair of fused silica wedges placed in the 1030 nm arm. The recombined beams were sent through a 1 mm thick calcite plate to ensure that their polarization were perfectly crossed. Finally, a 30 cm lens focused them into the interaction chamber of a VMI spectrometer. To compensate for the chromatic aberration of the lens, a telescope was placed in the 1030 nm arm of the interferometer, and was used to spatially overlap the focii of the $\omega$ and $2\omega$ beams. The VMI projected the photoelectron angular distribution onto a set of microchannel plates parallel to the  plane defined by the 1030 nm polarization $y$ and the laser propagation direction $z$. The MCPs were imaged by a phosphor screen and a SCMOS camera. 

Figure~\ref{FigXP} (a,g,m) shows the total photoelectron signal, measured by the VMI, at two different intensities of the fundamental laser field, with $r$ kept at $\sim 0.3$. As in the calculations, we observe a clear transition from a multiphoton ionization regime, having only few ATI peaks, to a strong-field ionization regime showing many ATI peaks and a narrow momentum distribution. As in the theoretical analysis, we extract the ESCARGOT signal by isolating the antisymmetric part of the signal with respect to $y,z$ (see Methods section). Controlling the two-color delay leads to a periodic modulation of the antisymmetric signal at twice the fundamental frequency. For each measurement, the delay was scanned over a range of 270 fs, corresponding to a $\sim 980$ rad scan in the relative phase between the two fields.  We Fourier transform the antisymmetric signal at each pixel on the detector, with respect to the two-color phase, and extract the $2\omega$ component, isolating the response of the ESCARGOT signal to every phase two-color phase. Panels (b-f),(h-l),(n-r) show the extracted $2\omega$ component of the ESCARGOT for (+)-Fenchone (b-f),(h-l) and for (+)-Camphor (n-r).

Our measurements reveals that, as predicted by the calculations, a significant ESCARGOT signal exists for all relative phases $\varphi$ between the two components of the electric field, and in all ionization regimes. In particular, we confirm that C-shaped electric field can produce significant forward/backward up/down asymmetries in chiral photoionization. Taking a closer look at the (+)-Fenchone measurement, performed in the multiphoton regime, (Figure~\ref{FigXP} (b-f)), one can observe sharp features in the low momentum range - up to about 0.2 a.u., and a smooth and relatively uniform signal at higher energies. In the low momentum region, the ESCARGOT signal maximizes to $\sim 0.5 \pm 0.1\%$ when $\varphi=0$ (``C-shaped field''), and is minimum when $\varphi=\pi/2$ (``8-shaped'' field).  We estimate the error bars by analyzing a 270 fs delay scan as 10 subscans of 27 fs, and calculating the 95$\%$ confidence interval by statistical analysis of the results. The situation is different in the higher momentum region, where the ESCARGOT is very low with the ``C-shaped'' field. As we go into the strong-field regime (h-l), the low momentum angular structures vanish so that the ESCARGOT signal seems more uniform. It also reaches lower values, maximizing in the $0.10 \pm 0.03\%$ range. This decrease is due to a lower influence of the chiral potential in the photoionization process as the laser field becomes stronger, as already observed in PECD experiments \cite{Beaulieu2016}. We also observe a noticeable difference between the asymetries recorded at $\phi=\pi/4$  and $3\pi/4$ as expected from the calculation. In order to check the enantiosensitive nature of these results, we repeated the measurements in racemic Fenchone (see Methods section). At low laser intensity, the ESCARGOT signal completely vanishes, which confirms the enantiosensitivity of the measurement. At high intensity, some non-zero components are present in the ESCARGOT signal, indicating the existence of artifacts in the experiment. However comparing the racemic and enantiopure ESCARGOT signals shows that the latter is dominated by the chiral-sensitive response. 

The experimental results from fenchone are in remarkable qualitative agreement with the TDSE simulations presented above, which may seem surprising given the simplicity of the molecular model compared to fenchone  and the difference in wavelength and pulse duration of the ionizing radiation. Repeating the measurements in another molecule (camphor, an isomer of fenchone) reveals that the agreement was coincidental. The characteristic sharp angular structures in the low momentum part of the ESCARGOT signal are not present in camphor (Figure~\ref{FigXP} (m-r)), and the overall asymmetry does not maximize for a C-shaped field, but in an intermediate configuration. This demonstrates that the ESCARGOT signal strongly depends not only on the laser intensity and two-color delay but also on the molecule under study. This sensitivity to both the molecular potential and the structure of the laser field reflect the physical origin of the ESCARGOT process -- the electron scattering driven by the chiral field in the chiral potential. It is thus important to investigate the timescale of these dynamics by analyzing electron trajectories.

\section{Semi-classical interpretation in the strong field regime}

\begin{figure}[!t]
	\begin{center}
		\centering{\includegraphics*[width=\columnwidth]{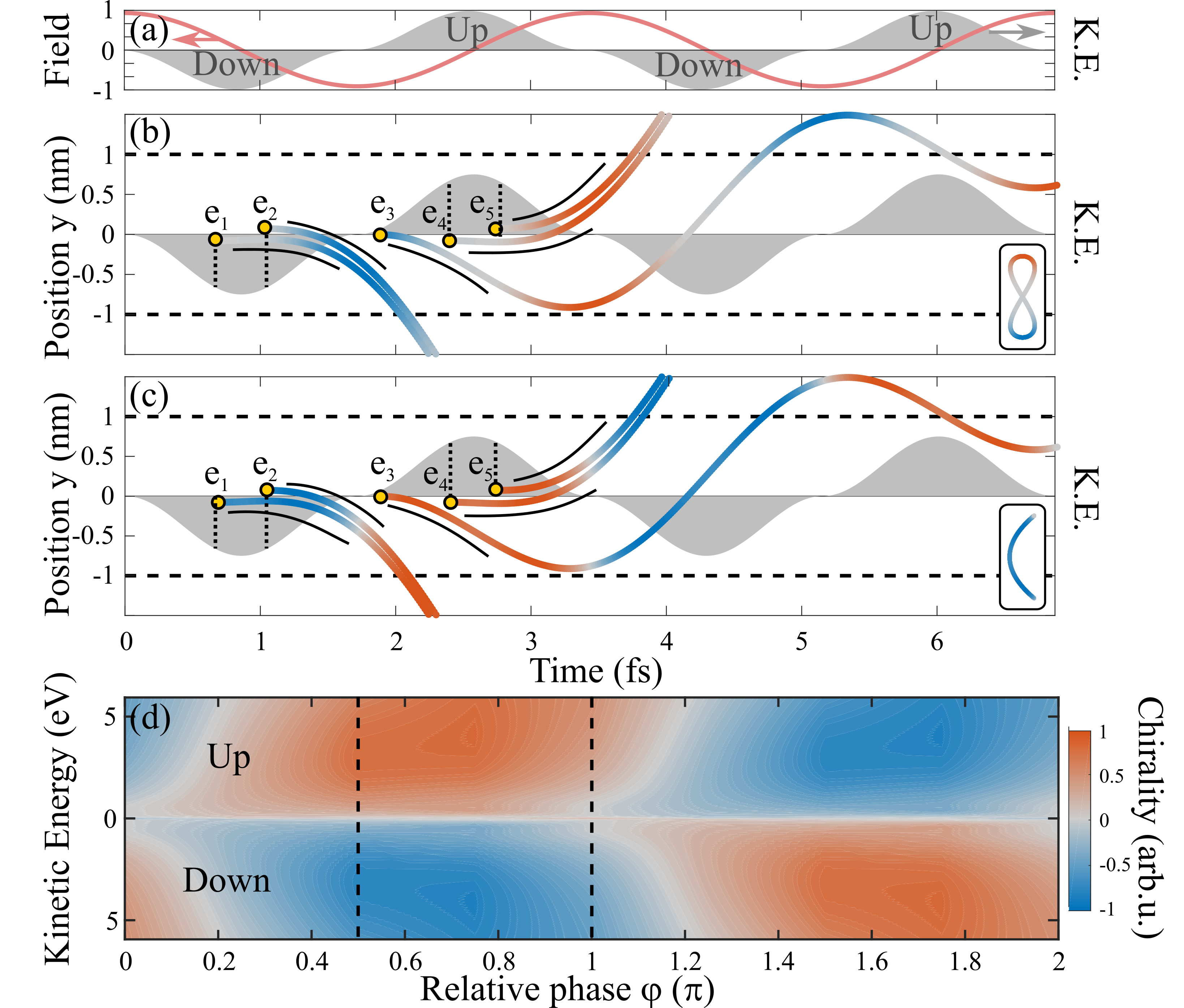}}
		\caption{Sub-cycle control over the instantaneous chirality of the field. (a) Oscillation of the fundamental electric field (red) and final kinetic energy of the electrons as a function of their ionization time (gray). The kinetic energy is plotted as a positive value for electrons with an upwards final velocity and as negative for electrons with a final downwards velocity, in order to differentiate between electrons reaching each half of the detector. (b-c) Two pairs of electron trajectories (e$_{1,2}$ and e$_{4,5}$), each pair leading to the same final kinetic energy, and a trajectory of a low energy electron (e$_3$). The trajectories are plotted for $\varphi=\nicefrac{\pi}{2}$ (b) and $\varphi=\pi$  (c). The colormap depicts the instantaneous chirality of the field. (d) Instantaneous chirality integrated over the time duration in which each electron is closer than 1nm to the molecular core, vs. two-color phase and final kinetic energy of the electron.
	}
		\label{fig2}
	\end{center}
\end{figure}
In order to have an intuitive picture of the photoionization dynamics underlying the ESCARGOT signal, we focus on the strong-field regime. The strong-field interaction defines a set of electron trajectories, providing a direct mapping between the ionization time and final momentum \cite{shafir12,Smirnova2013}. The forward-backward asymmetry in the photoionization of chiral molecules is imprinted during the electron scattering into the molecular potential. To estimate the influence of the instantaneous chirality on the photoionization process, it is thus necessary to evaluate the time spent by the electron in the vicinity of the molecular potential. We performed classical trajectory calculations within the strong-field approximation. The electrons are launched into the continuum with zero velocity and accelerated by the two-color laser field. The influence of the ionic core is neglected, such that these calculations do not include any molecular chirality and are thus very qualitative. 
Indeed, since our study focuses on the strong field regime, the electron trajectories in the (x,y) plane are dominated by the laser field, the effect of the molecular potential in this plane being a small correction. On the other hand, the molecular potential is the only force at play along the laser propagation direction (z), and gives rise to the chiral signal.

Figure \ref{fig2}(a) shows the kinetic energy acquired by ionized electrons as a function of their ionization time. Electrons born close to the maxima of the electric field end up with low energy, while electrons born close to the zeros of the field end up with higher energy. Figure \ref{fig2}(b-c) depicts a few typical electron trajectories, for two shapes of the electric field. The trajectories are very similar because they are driven by the strong fundamental component of the electric field. However the instantaneous chirality experienced by the electrons, represented as the colormap of the trajectory lines, strongly changes with the field shape. When the field has an 8 shape ($\varphi=\pi/2$, Figure~\ref{fig2}(b)), the electron 1, born at $t=0.2T_0$ (with $T_0=3.44$ fs, the periodicity of the 1030 nm field), experiences an almost zero optical chirality during a very short time, followed by a negative optical chirality from $0.25T_0$ to $0.65T_0$. We roughly estimate that the influence of the chiral potential is restricted to a distance of $\sim 1$ nm from the core. In this area the electron 1 has thus mainly experienced a negative optical chirality. Electron 2, born at $t=0.3T_0$, sees a negative chirality during all its travel in the first nanometer. It ends up on the same side of the detector ($y<0$) and with the same kinetic energy as electron 1. Thus, while these two electrons have not exactly experienced the same optical chirality and ionization dynamics in the two-color field, they have seen electric fields with the same helicity and their forward/backward asymmetries should add up. Electrons 4 and 5, born half a cycle later, have a symmetric behavior: they end up on the upper part of the detector, and mainly experience a positive optical chirality. The situation is less clear for electron 3, which is a low energy electron: it travels back and forth around the core, experiencing opposite optical chiralities. 

This analysis can be repeated in the case of a C-shaped field (Figure~\ref{fig2}(c)). Even if the C-shaped field rotates in the same direction in the upper and lower hemispheres, the instantaneous chiralities experienced by photoelectrons ending up in the upper and lower part of the detector are opposite. This simple analysis predicts that the accumulated optical chirality should be lower than with the 8-shaped field because it changes sign while the electrons are still in the vicinity of the ionic core. 

Neufeld and Cohen have introduced the concept of non-instantaneous chiroptical effects on a given timescale \cite{neufeld18}. We follow this approach to calculate the optical chirality accumulated over the timescale the electron takes to leave the chiral ionic core, by integrating the instantaneous chirality. In order to mimic the spatial extent of the chiral potential, time integration is further weighted by a gaussian profile centered on the origin with a 1 nm FWHM. Figure ~\ref{fig2}(d) shows the integrated chirality as a function of the electron kinetic energy and the relative phase $\varphi$. The results show that the chiralities experienced by upper and lower electrons are always opposite. The shape of the laser field that maximizes or minimizes the accumulated chirality depends on the kinetic energy of the electrons, explaining qualitatively why the ESCARGOT signal does not maximize at the same $\varphi$ for all electron energies. This model is however far too simple to quantify the ESCARGOT signal, which results from the scattering of the electrons in the chiral potential. What this analysis demonstrates is that shaping the vectorial electric field enables controlling the sub-cycle optical chirality experienced by the electrons, and that electrons detected in the upper and lower hemispheres see opposite optical chiralities for all shapes of the electric field.

\section{Conclusions and outlook}
The joint theoretical and experimental study presented here has enabled us to demonstrate that the instantaneous chirality of light could play a major role in chiral light-matter interaction: even fields with zero net chirality can induce sub-cycle chiroptical effects, which can be detected using appropriate differential detection schemes. The high sensitivity of the photoelectron angular distributions from chiral molecules to the instantaneous chirality of the ionizing light is the signature of the ultrafast nature of the photoionization process.  Recording the full 3D photoelectron momentum distribution, using direct delay lines detectors or a tomographic reconstruction, will resolve the sub-cycle temporal evolution of chiral photoionization.

The ESCARGOT technique is complementary to the other attosecond-resolved chiroptical methods that have emerged in the past few years. Chiral-sensitive high-order harmonic generation \cite{Cireasa15} probes the hole dynamics induced by a strong laser field with a resolution of a few tens of attoseconds. It is thus sensitive to attosecond ionic dynamics. Phase-resolved PECD \cite{BeaulieuScience2017} probes the difference between the photoionization delays of electrons ejected forward and backward with respect to the laser propagation axis, revealing the influence of the chiral molecular potential on the scattering dynamics of the outgoing electrons. ESCARGOT probes influence of the optical chirality on these scattering dynamics, and is thus complementary to phase-resolved PECD.

Sub-cycle shaped electric fields have led to many important achievements in attosecond spectroscopy \cite{Calegari16,Pedatzur15,Mairesse16}. However photoionization imaging experiments become extremely complicated as the molecules get larger, in particular because of orientation averaging effects. Chiral targets do not suffer from this -- chiroptical signals survive orientation averaging. We thus envisage that ESCARGOT will be a very powerful probe for photoionization imaging. In particular the detection of high energy rescattering electrons and their interference with direct electrons will enable unique holographic imaging of complex chiral molecules. The two-color scheme used here is only an example of sub-cycle polarization shaping. More sophisticated configurations, mixing linearly and elliptically polarized fields of different colors, could be used to increase the number of control knobs on the optical chirality gating.

\section*{Methods}

\subsection{TDSE calculations on toy model chiral system}

\subsubsection{Details on the theoretical framework}

The bound and continuum eigenstates $\phi_{E_i}$ of the toy model molecule described in the main text are obtained by diagonalizing the field-free Hamiltonian $H_0$ 
in a basis of functions $j_l(k_lr){\cal Y}^{sin,cos}_{lm}(\Omega_{\bf r})$ with $0 \le l \le l_{max}$ and $-l \le m \le l$. 
${\cal Y}^{sin,cos}$ are (sine and cosine) real spherical harmonics, and $j_l(k_lr)$ are spherical Bessel functions. In practice, we confine the electron motion 
within an hermetic spherical box of radius $r_{max}$ so that we only include in our basis the $j_l(k_lr)$ functions with $k_l$ such that $j_l(k_lr_{max})=0$. 
Electron momenta $k_l$ are further restricted from 0 to $k_{max}$. 
Such a procedure has led to a reliable representation of both bound and continuum states, and associated excitation and ionization processes, 
in atoms \cite{Pons2000} and diatomics \cite{Pons2004}. Here we mainly use $r_{max}=200$ a.u., $l_{max}=10$ and $k_{max}=7$ a.u. but convergence checks with 
respect to the values of $r_{max}$ and $l_{max}$ are presented below. We illustrate in Figure~\ref{Bern1} the molecular skeleton of the toy model system
as well as the shape of the fundamental bound state with ionization potential $I_P=8.98$ eV. 

\begin{figure}[!t]
\begin{center}
  \centering{\includegraphics*[width=0.5\columnwidth]{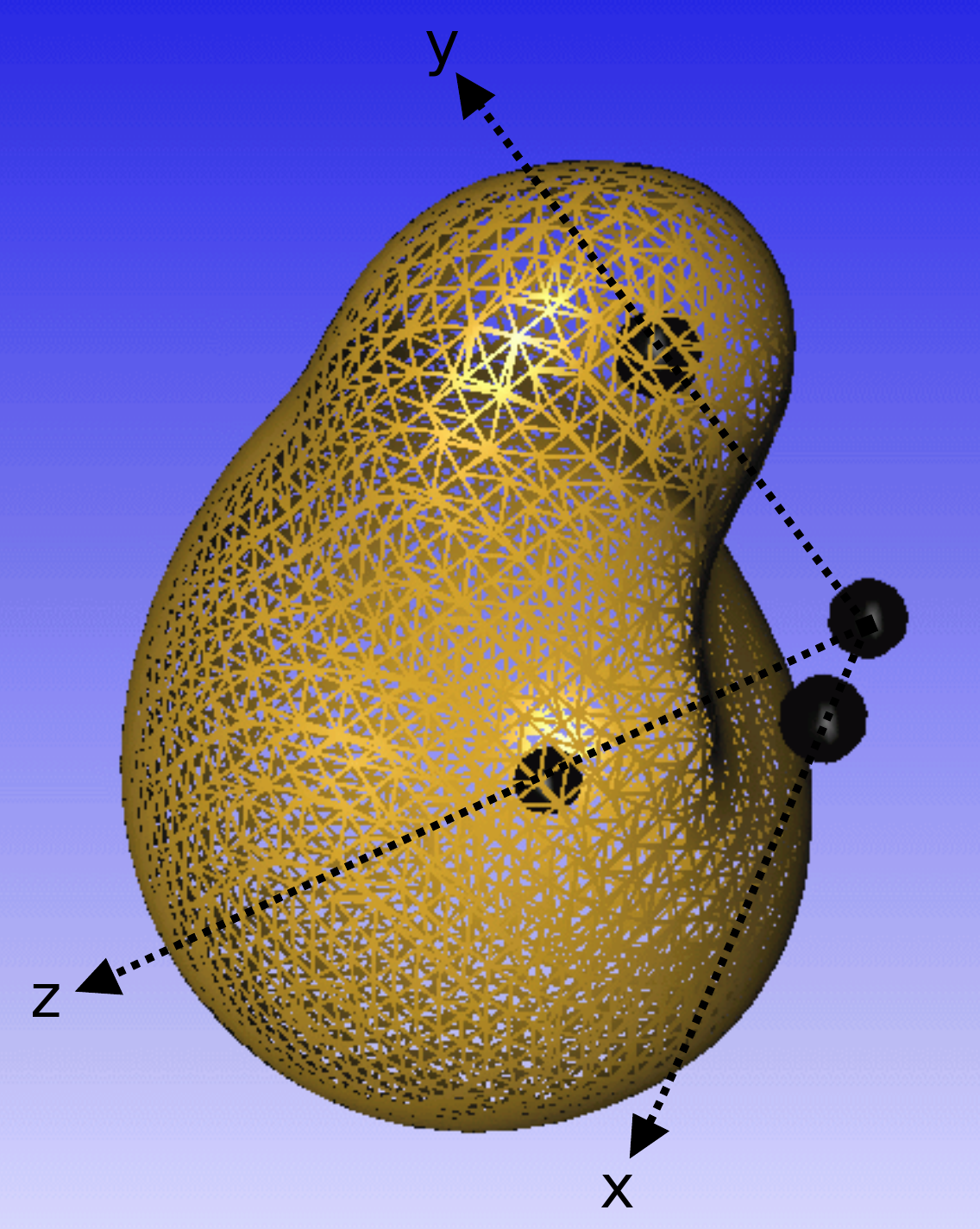}}
  \caption{Skeleton and isocontour representation of the fundamental state of the toy model chiral molecule employed in the TDSE calculations.}
  \label{Bern1}
\end{center}
\end{figure}

Once the eigenstates are obtained, we solve the TDSE by inserting the spectral decomposition 
$\Psi({\bf \hat{R}};{\bf r},t)=\sum_{i}{a_i({\bf \hat{R}};t) \phi_{E_i}({\bf r}) \exp(-i E_i t)}$ into eq. (\ref{TDSE}), yielding the system of
coupled differential equations for the expansion coefficients $a_i({\bf \hat{R}};t)$:
\begin{eqnarray}
\dot{a}_i({\bf \hat{R}};t)&=&-{\cal R}(\hat{{\text {\bf R}}})[{\bf A}(t)]\sum_{E_j}a_j({\bf \hat{R}};t) \times \nonumber \\
&<&\phi_{E_i}|{\bf \nabla} |\phi_{E_j}>{\text e}^{-i(E_j-E_i)t}
\end{eqnarray}
in velocity gauge, ${\cal R}(\hat{{\text {\bf R}}})[{\bf A}(t)]$ being the potential vector passively rotated to the molecular frame. 
This system is numerically solved for each molecular orientation ${\bf \hat{R}}$, subject to the initial conditions 
$a_i({\bf \hat{R}};0)=\delta_{i0}$ using the Cash-Karp (adaptative Runge-Kutta)
technique \cite{NumericalRecipes} after straightforward computation of the dipole couplings $<\phi_{E_i}|{\bf \nabla} |\phi_{E_j}>$. 

The ionizing part of the total wavefunction is defined by eq. (\ref{psi_ion}) in the molecular frame. It can be alternatively written as 
\begin{eqnarray}
\Psi^{ion}({\bf \hat{R}};{\bf r},t_f)=\sum_{l,m,j}b_{lmj}({\bf \hat{R}};t_f) \times \\
j_{l}(k_{lj}r){\cal Y}^{sin,cos}_{lm}(\Omega_{\bf r})  \nonumber
\end{eqnarray}
where
\begin{eqnarray}
b_{lmj}({\bf \hat{R}};t_f)=\sum_{l',m',i,E_i>0}{a_i({\bf \hat{R}};t_f){\text e}^{-i E_i t_f} }\times  \\
D_{il'm'j}\delta_{l',l}\delta_{m',m},\nonumber
\end{eqnarray}
$D_{il'm'j}$ being the diagonalization coefficients such that
$\phi_{E_i}({\bf r})=\sum_{l',m',j}{D_{il'm'j} j_{l'}(k_{l'j}r) {\cal Y}^{sin,cos}_{l'm'}(\Omega_{\bf r})}$.
The ionizing part of the total wavefunction can then be expressed in the lab frame according to
\begin{eqnarray}
\Psi^{ion}_{lab}({\bf \hat{R}};{\bf r},t_f)=\sum_{l,m,j}b_{lmj}({\bf \hat{R}};t_f) \times \\
j_{l}(k_{lj}r) {\cal R}^{-1}({\bf \hat{R}})\left[ {\cal Y}^{sin,cos}_{lm}(\Omega_{\bf r}) \right]  \nonumber
\end{eqnarray}
or equivalently, in momentum space
\begin{eqnarray}
\Psi^{ion}_{lab}({\bf \hat{R}};{\bf p},t_f)=\sum_{l,m,j}b_{lmj}({\bf \hat{R}};t_f)  \times \\
{\tilde j}_{lj}(p) {\cal R}^{-1}({\bf \hat{R}}) \left[ {\cal Y}^{sin,cos}_{lm}(\Omega_{\bf p}) \right]  \nonumber
\end{eqnarray}
where ${\tilde j}_{lj}(p)$ is the radial part of the momentum wavefunction associated to the confined $j_{l}(k_{lj}r){\cal Y}^{sin,cos}_{lm}(\Omega_{\bf r})$
primitive function \cite{Bransden}.

Note that the whole spectral TDSE scheme is the counterpart of the grid-TDSE approach presented e.g. in \cite{Demekhin2015,Baumert2018}. 
Further technical details about our multi-center spectral TDSE approach as well as its capabilities and limitations will be presented elsewhere.

\subsubsection{Influence of short pulse duration}
\begin{figure}[!t]
	\begin{center}
		\centering{\includegraphics[width=\columnwidth]{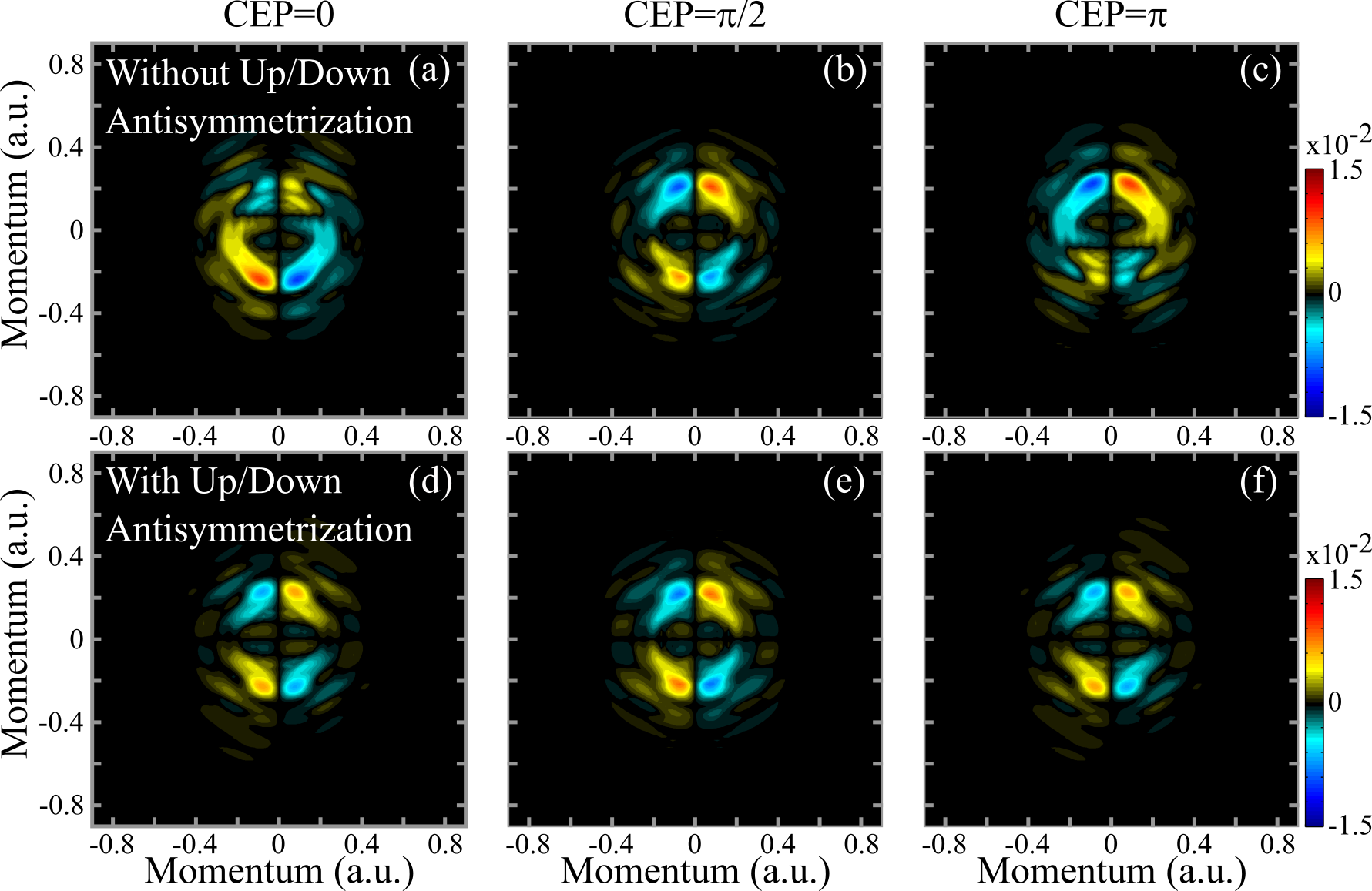}}
		\caption{Influence of the carrier-envelop phase $\phi_{CEP}$ in the TDSE calculations at $I=5\times 10^{12}$ W.cm$^{-2}$ and for a relative phase between the two components of the laser field $\varphi=3\pi/4$. (a-c) Projections of the ESCARGOT signal, defined as the antisymmetric part of the photoelectron distribution with respect to the propagation axis $z$. (d-f) Corresponding projections of the ESCARGOT signal, defined as the antisymmetric part of the photoelectron distribution with respect to the propagation axis $z$ and fundamental laser polarization axis $y$. The carrier-envelop phase is $\phi_{CEP}=0$ in (a,c), $\phi_{CEP}=\pi/2$ in (b,e), and $\phi_{CEP}=\pi$ in (c,f).
}
		\label{FigTHAnti}
	\end{center}
\end{figure}

The limited number of optical cycles of the laser pulses used in the calculations can lead to artifacts, related to carrier-envelop phase effects. 
In ($y,z$)-projections of the electron momentum distributions, these artifacts appear as asymmetries along the polarization direction $y$ of the composite laser field. In 
the theretical data presented in the paper we have therefore chosen to force both the symmetry of $\rho^{sym}_{VMI}(p_y,p_z)$ and antisymmetry of
$\rho^{anti}_{VMI}(p_y,p_z)$ along $p_y$. The impact of $p_y$-antisymmetrization of $\rho^{anti}_{VMI}$ is presented in Fig. \ref{FigTHAnti}, which compares the 
projections without (top) and with (bottom) the procedure, for three carrier-envelop phases of the laser. The calculations were performed using a $3\pi/4$ two-color phase, where the effect of the carrier-envelop phase was found to be maximum.  The figure shows that while carrier-envelop phase effects impact the raw ESCARGOT distributions, the up/down antisymmetrized distributions are independent of it. 

\subsubsection{Convergence checks}

\begin{figure}[!t]
	\begin{center}
		\centering{\includegraphics*[width=\columnwidth]{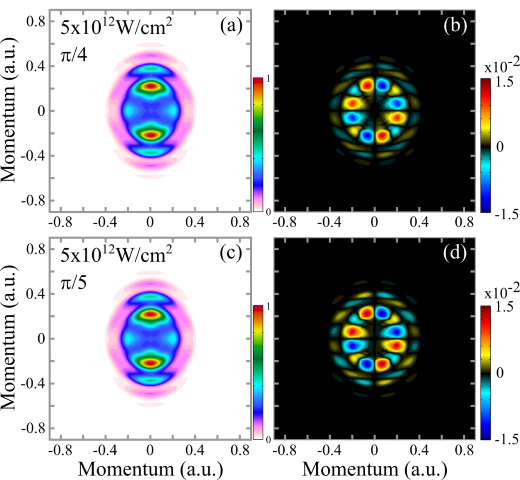}}
		\caption{ Convergence check with respect to the number of random molecular orientations included in the TDSE calculations: Symmetric (a) and antisymmetric (b) parts
			of the photoelectron momentum distribution obtained with an angular Euler spacing $\Delta \alpha = \Delta \beta = \Delta \gamma = \pi /4$. (c) and (d) are the counterparts of (a) and (b), respectively, obtained with $\Delta \alpha = \Delta \beta = \Delta \gamma = \pi /5$. All other TDSE parameters are the same in both calculations: $r_{max}$=200 a.u., $l_{max}=10$ and $k_{max}=7$ a.u. with two-color phase of $\varphi=0$. 
}
		\label{Bern2}
	\end{center}
\end{figure}

\begin{figure}[!t]
	\begin{center}
		\centering{\includegraphics*[width=\columnwidth]{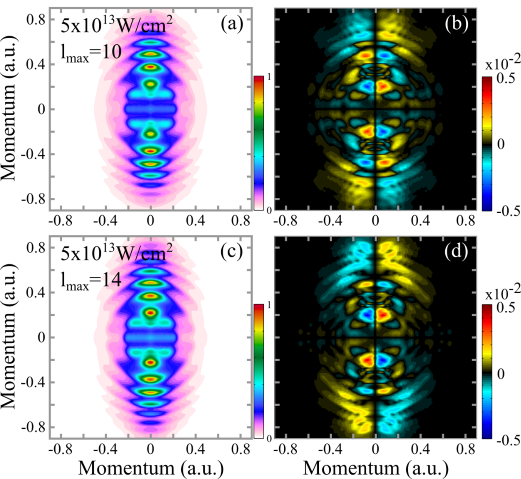}}
		\caption{Convergence check with respect to the maximum angular momentum included in the TDSE calculations: Symmetric (a) and antisymmetric (b) parts
			of the photoelectron momentum distribution obtained with $l_{max}=10$. (c) and (d) are the counterparts of (a) and (b), respectively, obtained with $l_{max}=14$. 
			All other TDSE parameters are the same in both calculations: $r_{max}$=200 a.u., $k_{max}=7$ a.u., $\Delta \alpha = \Delta \beta = \Delta \gamma = \pi /4$ with a two-color phase of $\varphi=0$. 
			}
		\label{Bern3}
	\end{center}
\end{figure}

It is stated in the main text that averaging over molecular orientations is performed in eq. (\ref{rho_p}) using numerical quadratures on
countable orientations with Euler angle spacing $\Delta \alpha = \Delta \beta = \Delta \gamma = \pi /4$. We illustrate in Figure~\ref{Bern2} how 
increasing the number of orientations according to $\Delta \alpha = \Delta \beta = \Delta \gamma = \pi /5$ left the symmetric and normalized 
antisymmetric (ESCARGOT) electron momentum distributions almost unchanged, for fundamental laser intensity I=5$\times$10$^{12}$ W/cm$^2$ 
and two-color delay $\varphi=0$. 

The other parameters of our numerical approach are $r_{max}$, $k_{max}$ and $l_{max}$. 

$r_{max}$ controls the extension of the
configuration space within which the electron dynamics can be fairly described. Too small $r_{max}$ value can lead to reflections of electron flux 
on the walls of the confinement spherical box and therefore results in blurred photoelectron momentum pictures. To minimize such effect, we
have considered rather short pulses (4 cycles of the fundamental 800 nm laser with 1 cycle ascending and descending ramps). Using $r_{max}$=200 a.u., 
we have indeed observed incoming waves resulting from reflection in our photoelectron momentum pictures, but they only alter the high-energy tail of the 
distributions, which are not of interest in the present work. In Figure~\ref{figTH} of the main text, small reflection patterns appear for the highest
laser intensity considered (I=5$\times$10$^{13}$ W/cm$^2$) and for electron momentum $p$ greater than 0.6 a.u. 

$k_{max}$ is the highest electron wavevector included in our basis of primitive Bessel functions. Diagonalization of $H_0$ therefore yields molecular eigenstates
whose energy extents up to $k_{max}^2/2$ (in a.u.). With $k_{max}=7$ a.u., extremely high-lying continuum states are then obtained. They are obviously not of interest
in the present investigation. However, large values of $k_{max}$ are necessary to obtain an accurate representation of the lowest lying eigenstates in terms
of (oscillating) Bessel functions. We accordingly increased $k_{max}$ up to the large value of 7 a.u. such that these lowest lying eigenstates, and especially the 
fundamental state, are fully converged in terms of eigenenergies and eigenfunctions. Subsequently to $H_0$ diagonalization, we only introduce in the dynamical
calculations the eigenstates whose energy is lower than 2 a.u. 

$l_{max}$ refers to both (i) the highest kinetic momentum available for single-center multipolar decomposition of any molecular (multi-centre) state and 
(ii) to the highest kinetic momenta the electron can gain through ionization. Multipolar decompositions of bound states converge quite rapidly so that 
(ii) is much more restrictive than (i), especially in the extremely non-linear multiphoton or strong-field regimes (see e.g. \cite{Cormier1996}). As stated in the main
text, we have mainly employed $l_{max}=10$. Focusing on the strong-field regime, where convergence with respect to increasing values of $l_{max}$ is the harder 
to reach, we show in Figure~\ref{Bern3} that increasing $l_{max}$ from 10 to 14 does not lead to significant changes in our photoelectron symmetric and
antisymmetric momentum distributions.

\subsection{Experimental data analysis}
The chiral response in the photoionization appears as a forward/backward asymmetry in the photoelectron spectra. This asymmetric part is quite small (typically below 1$\%$) and its detection can be challenging, because photoelectron angular distributions are never perfectly forward/backward symmetric even when using achiral targets, due to small spatial imperfection on the VMI detector. In PECD experiments this issue is typically circumvented by performing differential measurements, subtracting the spectra obtained using left and right circularly polarized radiation. This is not possible in the present scheme, because of the complex temporal structure of the electric field. To solve this problem we scan the relative phase between the $\omega-2\omega$ components of the electric field, and extract the ESCARGOT signal by Fourier analysis. 

\begin{figure}[!t]
	\begin{center}
		\centering{\includegraphics*[width=\columnwidth]{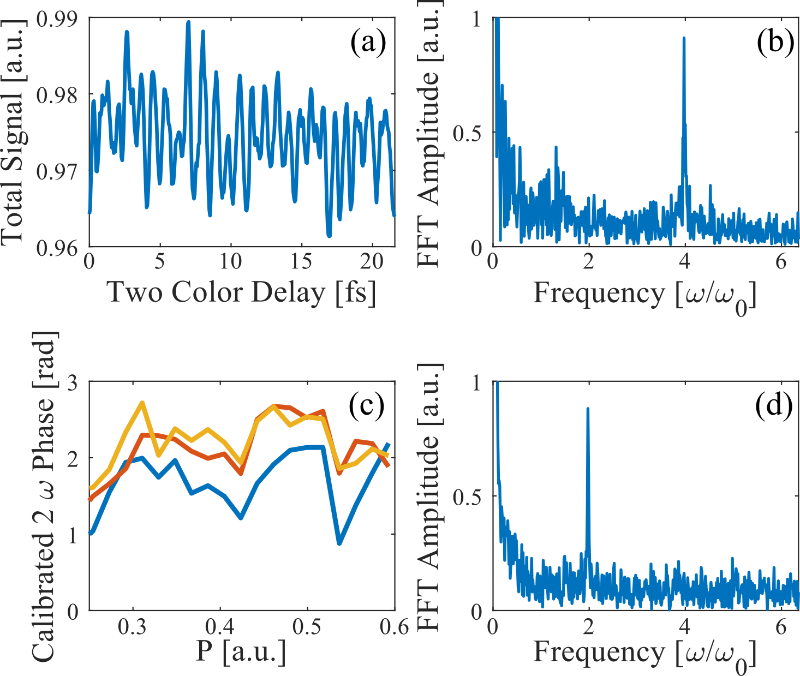}}
		\caption{{Calibration of the two-color phase using $4\omega$ oscillations. (a) Total photoelectron yield of a typical Fenchone(+) scan as a function of two-color delay, normalized to the maximal total signal in the scan. (b) Fourier transform amplitude of the signal in (a), in units of the fundamental frequency $\omega_0=291  THz$, the frequency of the 1030 nm field, showing a clear peak at the $4\omega$ frequency. (c) The phase of the antisymmetric signal as function of momentum P, for three Fenchone(+) scans. The phases were extracted by averaging over all the pixels with the same total final momentum in one quarter of the detector, for each image in the scan, and calculating the fourier transform of the time evolution of this averaged signal. Then the $2\omega$ phase of the oscillation was extracted, and calibrated according to the $4\omega$ oscillation phase of the total signal. The calibration is tested only for momentum higher than 0.25[a.u.] since in lower momenta, the asymmetric signal changes sign many times within each quarter of the detector, and therefore averaging over each quarter is meaningless. (d) An example for the fourier transform amplitude of the antisymmetric signal for momentum 0.4 [a.u.], in units of $\omega_0$.
		}}
		
		\label{Fig4wSupp}
	\end{center}
\end{figure}

As we scan the two-color delay, the photoelectron images shows oscillatory components. First, changes the field shape slightly modifies the ionization probability. The total yield is expected to maximize when the field amplitude is maximum, namely at $\varphi=0,\pi...$, and thus to oscillate at $4\omega$ frequency when scanning the delay. This appears clearly in Figure~\ref{Fig4wSupp} (a,b) which show the temporal evolution of the total yield for a typical (+)-Fenchone scan, and its Fourier transform. The phase of the $4\omega$ oscillating peak  provides a calibration of the two-color phase, up to a $\pi$ ambiguity. This means we have access to the shape of the electric field but not its direction of rotation. 

\begin{figure*}[!t]
	\begin{center}
		\centering{\includegraphics*[width=2\columnwidth]{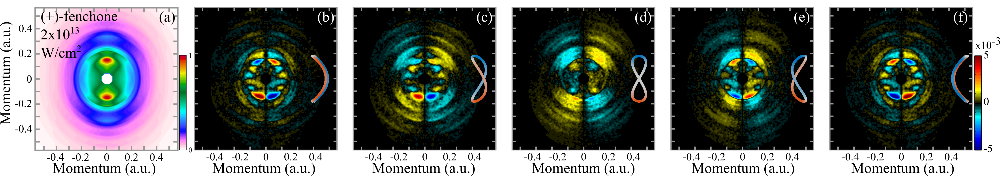}}
		\caption{Experimental data without up-down antisymmetrization. (a) Normalized projection in the y-z plane of the the photoelectron angular distribution for Fenchone(+) with a fundamental field of $2\times10^{13}$ W.cm$^{-2}$. (b-f) Evolution of the ESCARGOT signal as a function of the relative phase $\varphi$ between the two components of the ionizing field.  The shape of the two-color field, colored according to the instantaneous chirality, is shown next to each image. The ESCARGOT signal is normalized to the maximum of the photoelectron spectrum in (a).
		}
		
		\label{noAntiUpDown}
	\end{center}
\end{figure*}

By Fourier analyzing each pixel of the detector rather than the total yield, a slower modulation appears, at $2\omega$ frequency. Figure \ref{Fig4wSupp}(d) shows for instance the oscillating spectrum of one quarter of the detector (upper-forward). The $2\omega$ component is sensitive to the direction of rotation of the electric field, and is thus the chiral-sensitive signal. We forward/backward antisymmetrize the raw photoelectron images, Fourier transform them, and filter the $2\omega$ oscillating component in the frequency domain. The time evolution of the $2\omega$ antisymetric contribution is reconstructed from the amplitude and phase extracted with the Fourier analysis. The result of this procedure in the low-intensity ionization of fenchone is shown in Figure~\ref{noAntiUpDown}. The ESCARGOT signal shows quite good up/down antisymmetry, confirming that electrons ending in the upper and lower part of the detector have mostly experienced opposite optical chiralities, as seen in Figure~\ref{noAntiUpDown}. Residual up/down asymmetries exist, resulting from the imperfect nature of the detection. We thus antisymmetrize the ESCARGOT signals to eliminate these artifiacts.

\subsection{Comparison between (+)-Fenchone and a racemic mixture}
\begin{figure*}[!t]
	\begin{center}
		\centering{\includegraphics*[width=2\columnwidth]{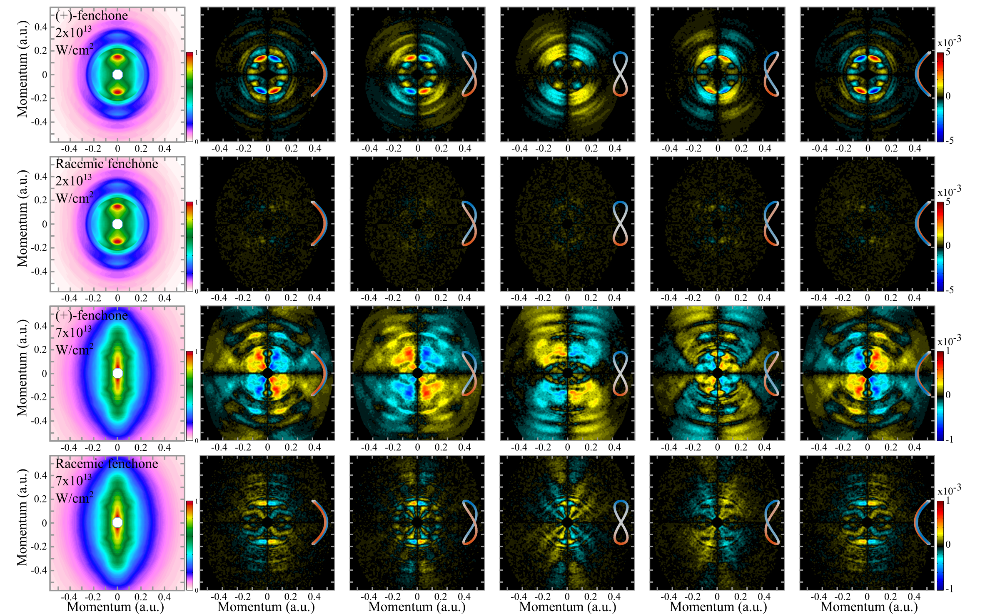}}
		\caption{Comparison of the ESCARGOT signal obtained in (+)-Fenchone (first and third rows) and a racemic mixture (second and forth rows), at $2\times10^{13}$ W.cm$^{-2}$ and $7\times10^{13}$ W.cm$^{-2}$. 
		}
		
		\label{RacemicFenchone}
	\end{center}
\end{figure*}

In order to validate that the ESCARGOT signal originates from the chiral properties of the molecular system, we performed a measurement in a racemic mixture of (+)-Fenchone and (-)-Fenchone. The mixture used was home made, mixing $55\%$ of a (-)-Fenchone sample with $82\%$ purity, and $45\%$ (+)-Fenchone sample with $100\%$ purity, giving a mixture with zero enantiomeric excess.  Figure~\ref{RacemicFenchone} shows a comparison between the ESCARGOT signals from (+)-Fenchone and the racemic mixture. At low intensity, the ESCARGOT signal almost completely vanishes in the racemic mixture. At higher laser intensity, some sharp non-zero features remain visible in the racemic mixture. This demonstrates that there are some residual artifacts in the experiment. This could be due for instance to imaging issues in the VMI detection because of the increase size of the photoelectron source at high intensity. However, most features observed in the enantiopure fenchone do disappear in the racemic mixture, and can thus safely be interpreted as a chiral signal.

\section*{Acknowledgements}

We thank R. Bouillaud, N. Fedorov and L. Merzeau for technical assistance. This project has received funding from the European Research Council (ERC) under the European Union's Horizon 2020 research and innovation program no. 682978 - EXCITERS, 336468 - MIDAS and from 654148 - LASERLAB-EUROPE. We acknowledge the financial support of the French National Research Agency through ANR-14-CE32-0014 MISFITS and from the R\'egion Nouvelle Aquitaine through RECHIRAM. 


%

\end{document}